# Cardiovascular risk and work stress in biomedical researchers in China: An observational, big data study protocol

Fang Zhu[1]*, Qian Zhang[1], Hao Chen,[2] Guocheng Shi[1], Chen Wen[1], Zhongqun Zhu[1], Huiwen Chen[1,3]

## Abstract

**Introduction:** Internet technologies could strengthen data collection and integration and have been used extensively in public health research. It is necessary to apply this technology to further investigate the behaviour and health of biomedical researchers. A browser-based extension was developed by researchers and clinicians to promote the collection and analysis of researchers' behavioural and psychological data. This protocol illustrates an observational study aimed at (1) characterising the health status of biomedical researchers in China and assessing work stress, job satisfaction, role conflict, role ambiguity, and family support; (2) identifying the association between work, behaviour, and health; and (3) investigating the association between behaviour and mental status. Our findings will contribute to the understanding of the influences of job, work environment, and family support on the mental and physical health of biomedical researchers.

**Methods and analysis:** This is a prospective observational study; all candidates will be recruited from China. Participants will install an extension on their Internet browsers, which will collect data when they are accessing PubMed. A web-based survey will be sent to the user interfaces every 6 months that will involve sociodemographic variables, perceived stress scale, job satisfaction scale, role conflict and ambiguity scale, and family support scale. Machine-learning algorithms will analyse the data generated during daily access.

**Ethics and dissemination:** This study received ethical approval from the ethics committee of the Shanghai Children's Medical Centre (reference number SCMCIRB-K2018082). Study results will be disseminated through peer-reviewed publications and conference presentations.

[1] Department of Cardiothoracic Surgery, Shanghai Children's Medical Center, Shanghai, China

[2] School of Medicine, Shanghai Jiao Tong University, Shanghai, China.

[3] Clinical Research Center, Shanghai Children's Medical Center, Shanghai, China.

* email: developer@scholarscope.cn

**Strengths and limitations of this study**

- The browser extension used in this study has become the most popular toolkit for literature reading.
- This nationwide study allows us to track participants' behaviours on the Internet, which provides objective, extensive, and accurate data for analysis.
- All the data are automatically collected by programmes, which reduces labour costs and eliminates the mistakes associated with data entry by investigators.
- The limitations of this study are: (1) we use a cross-sectional survey that may limit our ability to identify causal relationships between work stress, work environment, and family support, despite repeating the survey every half year, and (2) the approach involving self-reporting by researchers in an online survey may lead to response bias.

# Introduction

According to the data released by the United Nations Educational, Scientific and Cultural Organization (UNESCO), there were 8.7 million researchers (full-time equivalent) worldwide, of which 1.7 million were based in China.[1] The health and well-being of researchers is a significant topic in the area of public health. Over the last decade, interest in work stress, job satisfaction, and other relevant subjects concerning healthcare professionals has increased, with several cross-sectional studies reporting on the mental status of the participants.[2-6] However, to our knowledge, few reports have focused on the mental and physical health of biomedical researchers throughout China, which may be due to the large number and broad geographical distribution of this population.

There are more than 30 provincial administrative regions in China, most of which are in the central area and on the east coast. Meanwhile, the biomedical researcher population is also concentrated in these regions. Given the limitation of material and financial resources, many studies can only be performed in a single region or several regions. Moreover, unlike patients in medical centres, biomedical researchers do not gather in similar workplaces but instead are distributed among universities, hospitals, institutes, and companies. Therefore, big data collection and integration based on advanced Internet techniques is a solution to this heterogeneity. These techniques would promote our study of researchers' behaviour and psychology.[7]

Work-related stress is inevitable due to the demands of the contemporary work environment, which could incur low work efficiency and thus impose a considerable financial burden on the society.[8] Many maladies, including cardiovascular diseases,[9-11] have been linked to work stress in retrospective and prospective studies.[12, 13] However, the management of and research on work stress are often disregarded by policymakers and managers in China. According to a manuscript published in the *British Medical Journal*, Chinese researchers often work midnights and weekends. This situation is more severe than that in other countries,[6] and the overload of work in China has become a part of the academic culture.[14]

Many theories of stress have been presented, among which Lazarus and Folkman's cognitive theory of stress [15] is the most influential model; in this model, stress was defined as the interaction between the person and the environment. Many environmental factors could influence the cognition of work stress. In China, previous studies have suggested that policymakers should focus on improving job satisfaction to relieve stress in health workers.[16, 17] Wu *et al*. identified that role conflict and role ambiguity were crucial factors for occupational stress in doctors.[18] Lu *et al*. reported that work-family conflict and work stress are directly related to turnover intention.[19] Therefore, this study will mainly focus on the major factors influencing work stress and will provide an overview of biomedical researchers' cardiovascular and mental health. Big data analysis techniques will be used to explore the behaviour of the researchers.

## Methods and analysis

### Study design

This study is a prospective, observational study involving biomedical researchers in Mainland China since 4 July, 2018 (see the schematic in Figure 1). It is of great importance for these researchers to read scientific literature frequently. In China, PubMed is widely used by researchers because of its user-friendly interface and open resources. Therefore, we have developed a web browser extension that can collect the records when participants browse PubMed. Moreover, participants are invited to complete a questionnaire every half year via the browser extension, which collects demographic and work-related information.

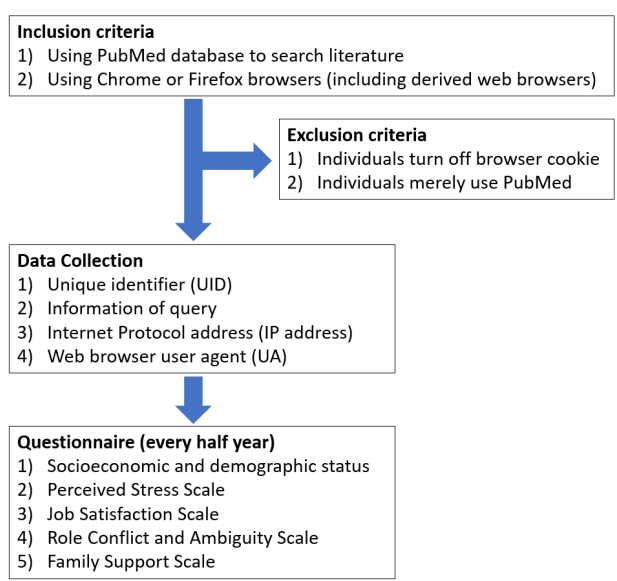

**Figure 1.** Schematic of the study design.

### Recruitment procedure

Participants will be recruited from the Internet. The browser extension can show related information, mark results, and share information with social communities when participants access PubMed. A website has been established to introduce this project and provide the extension for downloading (https://www.scholarscope.cn/). Participants who turn off browser cookies, or merely use PubMed (less than 7 days in total), will be excluded (Figure 1). All participants will receive an electronic informed consent form before they start using the extension.

The study participants will not be involved in the development of the research question or the study design. However, we have presented a feedback page, and useful suggestions will be integrated into this study.

## Data collection and pretreatment

### *Web browser extension*

The extension automatically collects data (Figure 2), including information about the query, Internet Protocol address (IP address), web browser user agent (UA), and unique identifier (UID).

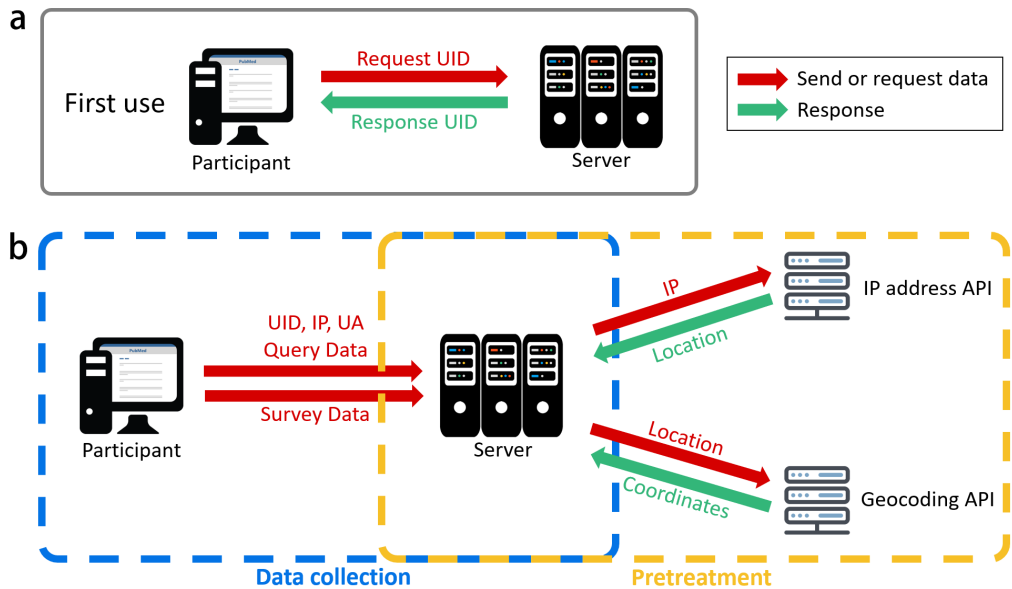

**Figure 2.** Schematic of data collection and pretreatment. a shows that when a participant first uses the extension, our server will allocate an UID to the participant. b illustrates the process of data collection and pretreatment. UID, unique identifier; IP, Internet Protocol address; UA, user agent; API, application programming interface.

UID is the only identifier used to distinguish participants. It will be calculated by the MD5 Message-Digest Algorithm[20] when participants first use the extension (Figure 2a). This sequence of 32 hexadecimal digits is stored in the browser's cookie. When the participant visits PubMed, the query information and UID will be recorded for further analysis.

Information regarding the query includes request dates, request time, keyword, and article number (PMID). Obtaining the information could help us analyse the reading habits and work schedules of participants. Meanwhile participants could be categorised by cross-matching with literature data in the PubMed database,[21] which may help us observe the differences in behaviour across different research areas.

We will adopt the big data analysis method combining the Internet techniques (Figure 2b). An IP address is a numerical label assigned to each device connected to the Internet. It is managed by the Internet Assigned Numbers Authority (IANA) and five regional Internet registries (RIRs). These organisations are responsible for IP address assignment in their jurisdictions. Therefore, the IP address could provide potential geographic information for each access. The Aliyun Application Programming Interface (API) [22] provides the function of converting IPs to city-level locations. However, due to international network environments and other reasons, participants often use a proxy or virtual private network (VPN) to access PubMed. To distinguish whether participants use an anonymous network connection, a classification method combining geolocation information will be used. Google Geocoding API can be used for converting addresses into geographic coordinates.[23] by using this API, the location could be converted to geographic coordinates. Considering that the speed of most civil airliners in the world is less than the speed of sound, if a participant moves to another remote location faster than the speed of sound, the IP address corresponding to this location will be labelled as an anonymous network node, and all information through this IP address would be classified into real geographic locations.

The web browser UA contains a characteristic string that allows network protocol peers to identify the application type, operating system, software vendor, or software version of the requesting software user agent. When a web browser sends a request, a header including the UA will be sent simultaneously and automatically.

*Assessment tools*

The study will consist of a self-administered questionnaire composed of five sections.

**Section 1 Basic Information:** This section mainly focuses on socioeconomic and demographic status and includes questions on age, body mass index, and education. This section also includes basic health information, such as systolic blood pressure, antihypertensive medication use, current smoking habits, and diabetes status. These variables can predict the long-term risk of cardiovascular disease according to the Framingham heart study.[24]

**Section 2 Perceived Stress Scale**: The Perceived Stress Scale (PSS), measuring the global stress that participants perceive in their life, has been widely used for measuring the perception of stress[25]. It is a five-point Likert type scale (0 = 'never' to 4 = 'very often') with 10 items, with a Cronbach's alpha of 0.78-0.86.[26-28]

**Section 3 Job Satisfaction Scale**: The job satisfaction scale (JSS) is a five-point Likert type scale (1 = very dissatisfied, 5 = very satisfied) with 15 items.[29] The Cronbach's alpha was 0.85-0.89 in Chinese populations.[30-32]

**Section 4 Role Conflict and Ambiguity Scale**: Role conflict and role ambiguity are assessed with the Rizzo scale.[33] Responses are based on a 5-point scale (1 = never, 5 = very often). Six items assess role ambiguity, and eight items assess role conflict. The Cronbach's alpha was 0.82 for role conflict and 0.80 for role ambiguity.[33]

**Section 5 Family Support Scale**: This scale was the family subscale of the multidimensional scale of perceived social support (MSPSS).[34] It is a seven-point Likert type scale (1 = very strongly disagree, 7 = very strongly agree) with four items; the Cronbach's alpha for this scale was 0.80-0.87.[34-36]

**Sample size and proposed statistical analysis**

Candidates meeting the criteria will be included. Based on the trial run, we estimate that this study will maintain more than 50 thousand participants and provide 1 gigabyte of data per week. Data will be reported in accordance with Strengthening the Reporting of Observational Studies in Epidemiology guidelines for observational studies.[37] All the data will be automatically collected by computer programs, and researchers will not interfere with this process. Considering the mechanism for collecting data, records with missing data are rare. Missingness may only occur when participants miss some questions in the questionnaires. If a participant does not complete the questionnaires or fills in an obviously wrong answer, this record will be manually deleted. The server needs to be maintained when necessary, during which data collection will be suspended. Missing data will be inputted based on an average of the data collected at the same time point in the previous week and the following week. For example, if server maintenance occurs at 6 pm on Friday, the average data on the previous Friday at 6 pm and the following Friday at 6 pm will be calculated and inputted as the lost part.

We will use R (version 3.5.3) for the statistical analysis. Normally distributed variables will be described by the mean and standard deviation. Differences between variables will be tested by the Student's t-test for continuous data and the $\chi 2$ test for categorical data. A p-value of 0.05 will be accepted as significant. For possible confounding variables, stratification or multivariable regression will be used. A multivariate, multinomial logistic regression model will be used to assess the association of risks and work stress.

To analyse the data generated by participants accessing PubMed, we will use multiple machine-learning methods. In supervised learning, a support vector machine (SVM) will be used to solve the binary classification problem while the k-nearest neighbour (k-NN) will be used to solve multiclass classification problems. In unsupervised learning, we will use the fuzzy c-means algorithm to solve the clustering problem.

**Privacy and data management**

All data will be automatically sent and stored by the programmes and servers. To control the quality of the questionnaire data, records with missing data will be deleted. The UID is the only label that distinguishes participants. Data will be stored on servers with firewalls and security measures. Access to records and data will be limited to study personnel.

**Ethics and dissemination**

This study received ethical approval from the ethics committee of the Shanghai Children's Medical Centre (reference number SCMCIRB-K2018082). Data and resources will be shared with other eligible investigators through academically-established means. The data sets analysed during the research will be available from the corresponding author upon reasonable request. The results from this work will be published in peer-reviewed manuscripts, presented at academic conferences, and summarized for the lay audience in report form.

**Strengths and limitations**

*Strengths*

The browser extension to be used in this study has become the most popular toolkit for literature reading. The extension gathers the data of biomedical researchers across China, enabling widespread collection and analysis of data. This study allows us to track participants' behaviours on the Internet, which provides extensive and accurate data for analysis. Furthermore, our study design, in which variables are repeatedly collected, will allow us to observe how factors change over time for each participant. Our programmes use advanced Internet techniques, such as browser cookies and the extension's local storage, which could greatly decrease the amount of lost data. All the data are automatically collected by programmes, which reduces labour costs and eliminates the mistakes associated with data entry by investigators. Therefore, the conclusions of this study would prove significant in China and may provide evidence for policy developments. It can also serve as a reference for other countries and regions.

*Limitations*

Our study has several limitations. The primary limitation is that we use a cross-sectional survey which may limit our ability to identify the causal relationships between work stress, work environment, and family support, although the survey will be repeated every 6 months. Meanwhile, self-reporting by researchers in an online survey may lead to response bias. In addition, we use the time of reading literature to evaluate the working hours of the participants. Though reading literature is an important part of research work, measurements based on the time for reading literature may underestimate working time.

**Authors' contributions**

Fang Zhu, Qian Zhang, and Huiwen Chen designed the study described in this protocol; Fang Zhu and Hao Chen wrote the programmes of web browser extension; Fang Zhu built and set up the web servers; Wen Chen and Guocheng Shi tested the robustness of the programme; Fang Zhu wrote the manuscript; Zhongqun Zhu and Huiwen Chen edited the manuscript drafts.

**Funding statement**

This research received no specific grant from any funding agency in the public, commercial, or not-for-profit sectors.

**Competing interests**

The authors declare no competing interests.